\documentclass[amsmath,amssymb,10pt,nofootinbib,superscriptaddress]{revtex4}
\usepackage{ graphicx, float,amsmath, amsmath, amssymb}
\usepackage[export]{adjustbox}
\usepackage[labelsep=period]{caption}
\usepackage[normalem]{ulem}
\usepackage{mathtools}
\usepackage{siunitx}
\usepackage{soul}
\usepackage{physics}
\usepackage{minitoc}

\usepackage{bm}
\usepackage{xcolor}
\usepackage{subfigure}

\definecolor{alizarin}{rgb}{0.82, 0.1, 0.26}

\def\be{\begin{equation}}
\def\ee{\end{equation}}
\def\bea{\begin{eqnarray}}
\def\eea{\end{eqnarray}}
\def\bse{\begin{subequations}}
\def\ese{\end{subequations}}
\def\Mc{{\cal M}_c}
\def\Vcav{V_{\text{cav}}}
\def\teff{t_{\rm eff}}

\begin{document}
\title{Prospects for detection of ultra high frequency gravitational waves \\ 
from compact binary coalescenses with resonant cavities}


\author{Aur\'{e}lien Barrau}
\affiliation{
Laboratoire de Physique Subatomique et de Cosmologie, Universit\'e Grenoble-Alpes, CNRS/IN2P3\\
53, avenue des Martyrs, 38026 Grenoble cedex, France
}

\author{Juan Garc\'ia-Bellido}
\affiliation{Instituto de F\'isica Te\'orica UAM/CSIC, Universidad Aut\'onoma de Madrid, Cantoblanco 28049 Madrid, Spain}

\author{Thierry Grenet}
\affiliation{
Institut Néel, Universit\'e Grenoble-Alpes, Grenoble INP, CNRS\\
25 avenue des Martyrs,
BP 166,
38042 Grenoble cedex 9
}

\author{Killian Martineau}
\affiliation{
Laboratoire de Physique Subatomique et de Cosmologie, Universit\'e Grenoble-Alpes, CNRS/IN2P3\\
53, avenue des Martyrs, 38026 Grenoble cedex, France
}

\date{\today}
\begin{abstract} 
This article aims at clarifying the situation about astrophysical sources that might be observed with haloscope experiments sensitive to gravitational waves in the 1-10 GHz band. The GrAHal setup is taken as a benchmark. We follow a very pedagogical path so that the full analysis can easily be used by the entire community who might not be familiar with the theoretical framework. Different relevant physical regimes are considered in details and some approximations encountered in the literature are questioned. In particular, we carefully take into account the fast drift of the gravitational wave frequency and the associated experimental limitations of different kinds. This has strong impact on sensitivity estimates. We also relax the usual assumption that only the merging phase should be considered. The distances that can be probed are carefully evaluated, 
taking into account degeneracies between physical parameters. 
We conclude that any detection in the near future is probably impossible. 
\end{abstract}
\maketitle

\section{Introduction}

It is quite extraordinary that gravitational waves above the MHz might, in principle, be observed thanks to resonant cavities \cite{Goryachev:2021zzn, PhysRevD.90.102005, Berlin:2021txa, Berlin:2023grv, Domcke:2022rgu, Domcke:2020yzq, PhysRevD.104.023524}.
This article aims at investigating in details the possible inspiralling binary mergers, taking the GrAHal experiment \cite{Grenet:2021vbb, GrenetFIPs22} as a benchmark. We follow a very pedagogical path so that all arguments are made clear to non-specialists. An excellent review of the possible sources can be found in \cite{Aggarwal:2020olq} whereas an in-depth investigation of the expected signal is given in \cite{Berlin:2021txa}. The search for light black holes was, in particular, considered in \cite{Herman:2020wao,Franciolini:2022htd}.

The basic idea behind the phenomenon is that a gravitational wave propagating through a static electromagnetic field sources a feeble electromagnetic wave at the frequency of the gravitational wave that might be seen by resonant detectors. Technically, the theoretical framework is well established and is the one of the so-called inverse Gertsenshtein effect \cite{Boccaletti:1969aj,Fuzfa:2015oaa,Fuzfa:2017ana,Ejlli:2019bqj}.\\

In this article, we consider binary systems of (light) black holes and define carefully the relevant variables. We do not focus on the {\it a priori} most favored region of the parameter space -- non is actually really favored -- but, the other way round, we try to investigate the entire range of mergers that might, in principle, be discovered. For most of the work, we remain agnostic about the formation mechanisms. We show that a huge range of masses is {\it a priori} visible for two reasons. First, because fixing the chirp mass, which enters the evaluation of the strain, does not fix the mass of each object. Second -- and more importantly -- because, for a given observation frequency, varying the amount time before coalescence at which the system is observed changes the relevant masses. Although neglected in most studies, this point is important because the actual mass of an hypothetical (light) binary system existing close to the Earth is {\it not} known. It could very well be different from the value corresponding to the best experimental sensitivity. Focusing on the latter case only  would be relevant  if the mass spectrum was known, wide, and flat -- thus ensuring that the ``best" situation inevitably occurs. This is not true for the kind of studies performed here: we do not know which black hole masses (if any) do exist in our cosmological neighborhood, hence the importance of considering the most general cases. As in particle phenomenology, one cannot consider the sensitivity of a detector to new particles only at the mass where the signal-to-noise ratio is the best but also at other values that might be the ones realized by Nature, even though they lead to a weaker signal.\\

After discussing in details the range of masses of interest, we investigate the maximum distance at which a specific source can be detected. This obviously depends not only on the astrophysical parameters but also on the details of the experimental setup. As in \cite{Franciolini:2022htd} and \cite{Domcke:2023bat},  we insist on introducing different important timescales and on defining relevant regimes that are sometimes overlooked in the usual literature although they play a fundamental role. Both the required strain and the maximal distance for detection are carefully studied. Our results differ from the formulas sometimes encountered in similar works, not only in numerical estimates but also in the functional dependence upon physical parameters. This is due to our ``integrated" treatment of the full process -- from the source to the detector --, making our output directly usable, evading the risk of evaluating things in a somewhat inconsistent way. 

For the sake of completeness we also consider less compact systems. This obviously makes the situation even worse. We take this opportunity to underline a common misunderstanding associated with compactness.\\ 

It should be underlined that quite similar questions were addressed in the pioneering work \cite{Franciolini:2022htd}. In particular, the emphasis was put on a state-of-the-art description of merger rates. It was concluded that detecting individual mergers with current proposals remains difficult. Our work is however very different as its two main ingredients were not considered in \cite{Franciolini:2022htd}. First, in the present article, we do not focus on the merging only but we investigate the entire parameter space opened by considering earlier stages of the coalescence. This changes the relevant masses at a given frequency by more that 10 orders of magnitude. This is important in itself as the actual masses are unknown but this becomes even more relevant as we show that the higher masses (corresponding to systems seen at merging) are {\it not} leading to a better sensitivity. Second, we take into account the temporal aspects more accurately by considering several regimes that are ignored in \cite{Franciolini:2022htd}. This is not a minor correction and this leads to an estimated sensitivity worsened by several orders of magnitude, therefore drastically changing the expected detection rate, even for systems observed at merging.\\

Numerical estimates are performed based on a realistic experiment -- the Grenoble Axion Haloscope (GrAHal). The GrAHal platform \cite{Grenet:2021vbb}
is mostly designed to search for axion dark matter, but was also believed to be an interesting gravitational wave detector in the ultra high frequency range. It relies  on a synergy of know-hows on key aspects of such experiments: high magnetic fields and fluxes, ultra-low temperatures, and very low noise quantum detectors. 

Although the expected sensitivity on the gravitational strain is quite impressive, this study shows that it remains insufficient for realistic astrophysical situations and that the orders of magnitude of distances to mergers that can be probed are very small. 
We conclude that binary systems of black holes are out of reach for cavities.

\section{Observable black hole masses}

When dealing with extremely high frequency gravitational waves, light black holes are obvious candidates. A binary system of stellar mass black holes would merge long before the frequency of the emitted gravitational waves reaches the band of interest for this work. Only low-mass black holes, presumably formed in the early universe, might lead to the desired signal.\\

Usually, the typical (total) mass $M$ of a system of black holes associated with a given experimental frequency $\nu$ is deduced from the innermost stable circular orbit, following from $\nu \sim f_{\rm ISCO}\propto M^{-1}$. This is reasonable in the sense that this corresponds to a signal with maximum strain. This, however, obviously not only discards a huge portion of the parameter space but also neglects the fact that the system spends a tiny amount of time within the detector bandwidth close to its merging whereas it can generate a long, coherent, signal earlier in the inspiralling process. This is taken into account in this work.\\

In the following, we assume non-spinning black holes and use the basic circular orbit approximation to fix orders of magnitude. If $m_1$ and $m_2$ are the masses of each black hole, the chirp mass is defined by:
\begin{equation}
\Mc=\frac{(m_1m_2)^{\frac{3}{5}}}{(m_1+m_2)^{\frac{1}{5}}}.
\end{equation}
The frequency of gravitational waves emitted at time $\tau$ before the merging is given by \cite{Maggiore:2007ulw}:
\begin{equation}
f(\tau)=\frac{1}{\pi} \left( \frac{5}{256} \frac{1}{\tau} \right)^{\frac{3}{8}} \left( \frac {G \Mc}{c^3} \right)^{-\frac{5}{8}}.
\label{freq}
\end{equation}
It is immediately clear that fixing $f(\tau)=\nu$ does {\it not} fix the mass $\Mc$ (the fact that choosing the frequency does not determine the mass non-ambiguously is actually trivial from Kepler's law). The frequency depends on the product $\tau^3\Mc^5$. When the time $\tau$ is increased, systems with smaller masses can lead to gravitational waves at the very same frequency. The price to pay is obviously a decrease in the amplitude of the signal, as the relevant masses are smaller and as the process has to be seen earlier in its evolution. However, the frequency of the system is then drifting more slowly and the relevant integrating time, hence the strain sensitivity, will substantially increase. In addition, as we shall quantify this later, a given chirp mass corresponds to a wide range of masses $m_1$ and $m_2$. It is therefore clear that even at a fixed observation frequency $\nu$, a wide range of masses can, in principle, be probed. 

More importantly, we would like to argue once more that the problem is {\it not} an optimization one, unless primordial black holes have a very extended mass spectrum, leading to a signal spread over a wide frequency range~\cite{Carr:2019kxo}. We simply do not know where the sources (if any) are and the entire parameter range should therefore be investigated, even if the sensitivity is lower in some regions. There is no point focusing on specific masses or on the physical case maximising the signal as this might very well not correspond to the parameters describing the real World. One should keep in mind that we deal here with exotic objects whose existence and characteristics are only predicted in very specific models. If real actual masses are such that the system frequency crosses the bandwidth of the experiment long before the merging, this situation has to be investigated as the emission at merging would anyway not be accessible to the experiment, whatever its intensity.

To go further in the characterisation of the range of accessible black hole masses we need to specify some experimental quantities.  The formulas given in this article are generic and can be used for a wide range of apparatus. We have however chosen to also express them as normalized to  fiducial values corresponding to the GrAHal experiment, as described in Table \ref{Table GrAHal configurations}. {In this table $T_{\text{sys}}$ is a benchmark for the system temperature corresponding to the usual sum of the cavity thermal noise $h \nu \left(\text{exp}(h \nu / k_B T) -1 \right)^{-1}$, $T$ being the cavity temperature, of the zero-point fluctuations of the blackbody gas and of the amplifier noise temperature (the dominant term).} Thanks to the LNCMI multiconfiguration hybrid magnet \cite{PugnatHybrid43T2022} the GrAHal project
will be able to cover a very wide frequency range for haloscopes. The extreme cases considered here, namely 0.34 GHz and 11.47 GHz, should therefore be enough to deal with all experiments of this type. Basically, the idea is to consider the search for ultra high frequencies gravitational waves as an additional feature of haloscopes, reusing the experimental setups and data. The cavity resonant mode of interest for the search of axions or Axion Like Particles (ALPs) being the $TM_{010}$ mode, the relevant frequency of the experimental setup will from now on be set to $\nu = \nu_{TM010}$. Different values of this frequency for the GrAHal platform, corresponding to different possible setups, are listed in Table \ref{Table GrAHal configurations}. As discussed in \cite{Berlin:2021txa}, due to the tensor nature of gravitational waves, it would be judicious to investigate the coupling with other resonant modes in the cavity, \textit{e.g.} with modes which exhibit a quadrupolar structure. However, as it will be discussed later in this article, an improvement of that kind would unfortunately not change the main conclusions. Hence, we stick to the usual $TM_{010}$ mode for this analysis, which allows a direct reuse of axionic data and facilitates comparisons with already existing studies. \\

\begin{table}[h!]
\begin{center}
\begin{tabular}{|c|c|c|c|  }  
 \hline
 \multicolumn{4}{|c|}{GrAHal configurations} \\
 \hline
$B_{\text{max}} (\text{T})$ & Cavity volume ($\text{m}^3)$ & Benchmark system noise temperature $T_{\text{sys}}$ $(\text{K})$ & $\nu_{TM010} (\text{GHz})$ \\
 \hline
9   & $5.01\times10^{-1}$    & 0.3 &   0.34 \\
 17.5 &   $3.22\times10^{-2}$   & 0.3 & 0.79 \\
 27 & $1.83\times10^{-3}$  & 0.4 &  2.67 \\
 40    & $1.42 \times10^{-4}$  & 1.0 &  6.74 \\
 43 &   $4.93 \times10^{-5}$   & 1.0 & 11.47 \\
 \hline
\end{tabular}
\caption{Main possible configurations for GrAHal using the LNCMI hybrid magnet \cite{PugnatHybrid43T2022}.}
\label{Table GrAHal configurations}
\end{center}
\end{table}

\begin{figure}[!h]
\null\hfill
\subfigure{\includegraphics[width=0.49\textwidth]{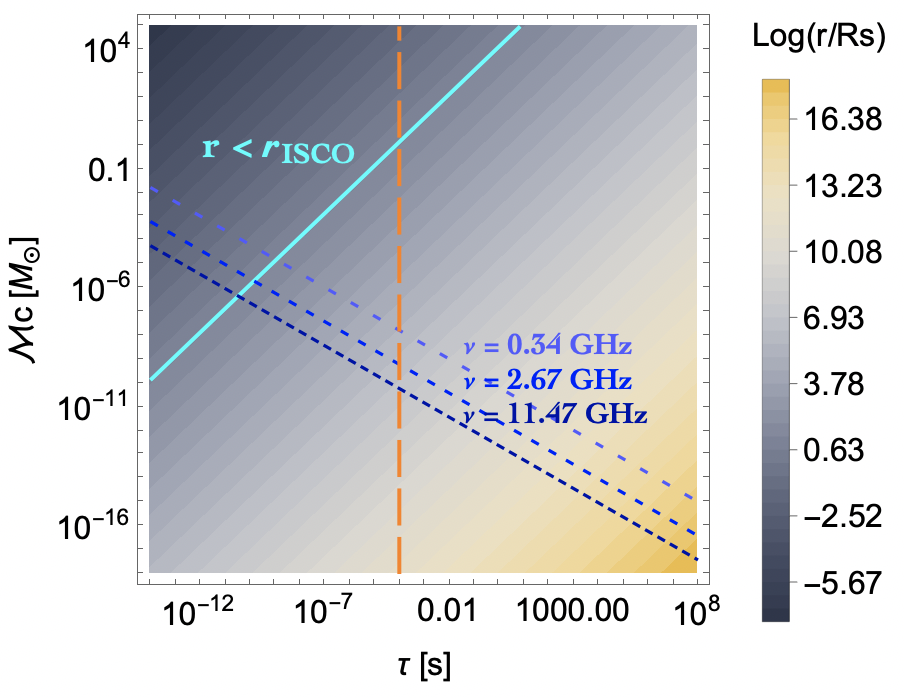}}
\hfill
\subfigure{
    \includegraphics[width=0.49\textwidth]{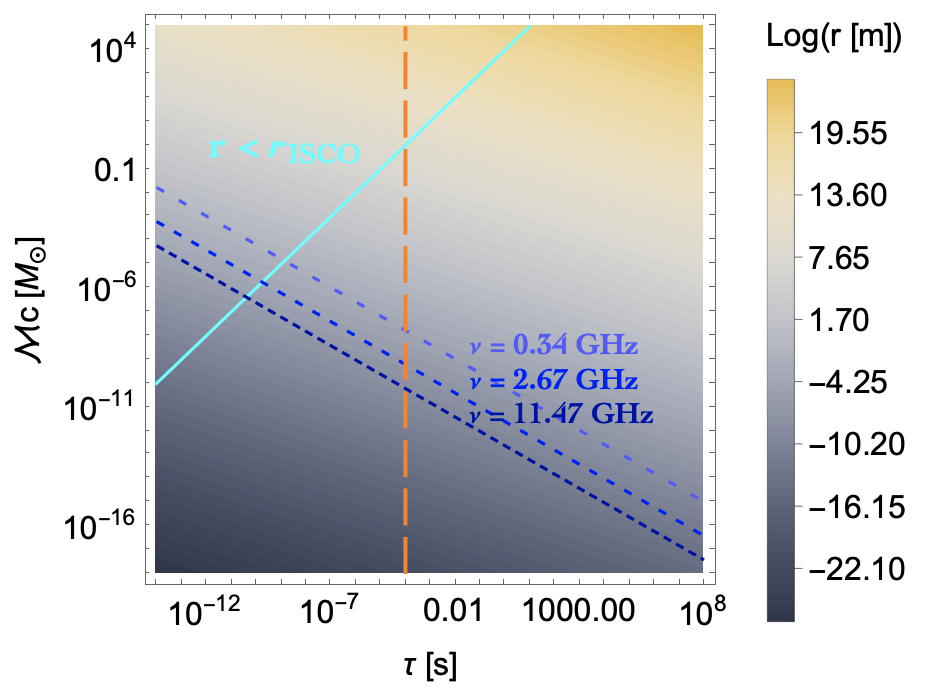}}
    \caption{\underline{Left panel}: Chirp mass of a binary system of black holes, in solar mass units, as a function of the time before merging for different detector frequencies. The background colors are obtained setting $m_1 = m_2$ and indicate the radial distance between the two bodies in units of their Schwarzschild radius. The slanted cyan line corresponds to this distance being equal to the ISCO radius whereas the vertical orange dashed line corresponds to $\tau = \Delta \nu^{-1}$. \underline{Right panel}: Same figure with the radial distance expressed in meters.}  
    \label{fig1}
\end{figure}

In Fig.\ref{fig1}, the chirp mass is displayed as a function of the time before merging for three of the previous configurations (the two extreme ones and the central one). For a given gravitational waves emission frequency -- which means a given orbital frequency -- there is a full degeneracy (ignoring for the moment other characteristics of the waves) between $\Mc$ and $\tau$. The chirp mass ranges from $10^{15} \text{g} \approx 10^{-18} M_\odot$ to $2 \times 10^{-5} M_\odot$ for the $\nu = 0.34$ GHz configuration. The upper limit becomes $2 \times 10^{-6} M_\odot$ for the $\nu = 2.67$ GHz, and $5 \times 10^{-7} M_\odot$ for the $\nu = 11.47$ GHz configuration. This covers more than 10 orders of magnitude in chirp masses.


For the unfamiliar reader, it might be worth clarifying a possible confusion. From Kepler's law it is obvious that considering higher masses at a given frequency requires to increase the distance between the black holes. Therefore it seems that they should be considered earlier in the inspiralling process, meaning that $\tau$ should be higher,  conflicting with Fig.\ref{fig1} and Eq.(\ref{freq}) which shows, the other way round, that $\tau$ is smaller for higher masses. The reason is simply that the distance depends not only on $\tau$ but also on $\Mc$. Otherwise stated, the heavier black holes are indeed at a larger distance -- still for a fixed frequency -- one from the other but they actually are {\it closer} to their merging time ($M \propto r^3$ which, using Eq.(\ref{Eq.RvsTau}), leads to $\tau \sim r^{-5}$). At fixed frequency, slightly counter-intuitively, the larger the distance, the smaller the merging time. 

The lower bound $M_*\approx 10^{15}$ g simply comes from the evaporation of light black holes through the Hawking mechanism \cite{Hawking:1974sw}. A black hole with mass $m<M_*$ would have fully evaporated in a time smaller than the age of the Universe. As we are confined to nearby objects, there is no chance to observe far-away lighter black holes. This is a firm limit if we assume that they were formed in the early universe. It might nevertheless make sense to also consider black holes at the Planck mass (that is 20 orders of magnitude less massive than the previously given lower bound) as quite a lot of scenarios of quantum gravity (see references in \cite{Barrau:2022eag}) predict that the evaporation ends with stable relics with masses $m_{rel}\approx M_{Pl}\approx 10^{-5}$ g. This would be an additional point in the plot of Fig. \ref{fig1} with $\tau \approx 10^{44}$ s (way higher than the age of the universe), for the 2.67 GHz configuration. Obviously, the strain for such gravitational waves is tiny beyond words and this is a purely mathematical situation.

The upper bound corresponds to a pair of black holes reaching their innermost stable circular orbit (ISCO). Basically, it means that the merging is actually taking place: this is the end of the game for the considered system and a single black hole is about to be formed. This limit can be easily evaluated by using the radial distance as a function of the time $\tau$ before merging for a quasi-circular orbit~\cite{Maggiore:2007ulw}:
\begin{equation}
r(\tau)=\left( \frac{256}{5} \frac{G^3M^2\mu}{c^5} \tau  \right)^{\frac{1}{4}},
\label{Eq.RvsTau}
\end{equation}
with $M=m_1+m_2$ and $\mu= m_1m_2/(m_1+m_2)$. Importantly, it should be noticed that $r$ does not depend directly on the chirp mass. However, it is impossible to observe black holes heavier than the mass corresponding to $m_1=m_2=2^{1/5}\Mc^{max}$, where $\Mc^{max}$ is the highest possible chirp mass corresponding to the chosen frequency. It is so because taking a value of $m_1$ different from $m_2$, so that $\Mc$ remains constant, inevitably increases $M$. As $r_{\rm ISCO}=6GM/c^2$, the system would have merged.\\

\begin{figure}
    \centering
    \includegraphics[width=0.67\textwidth]{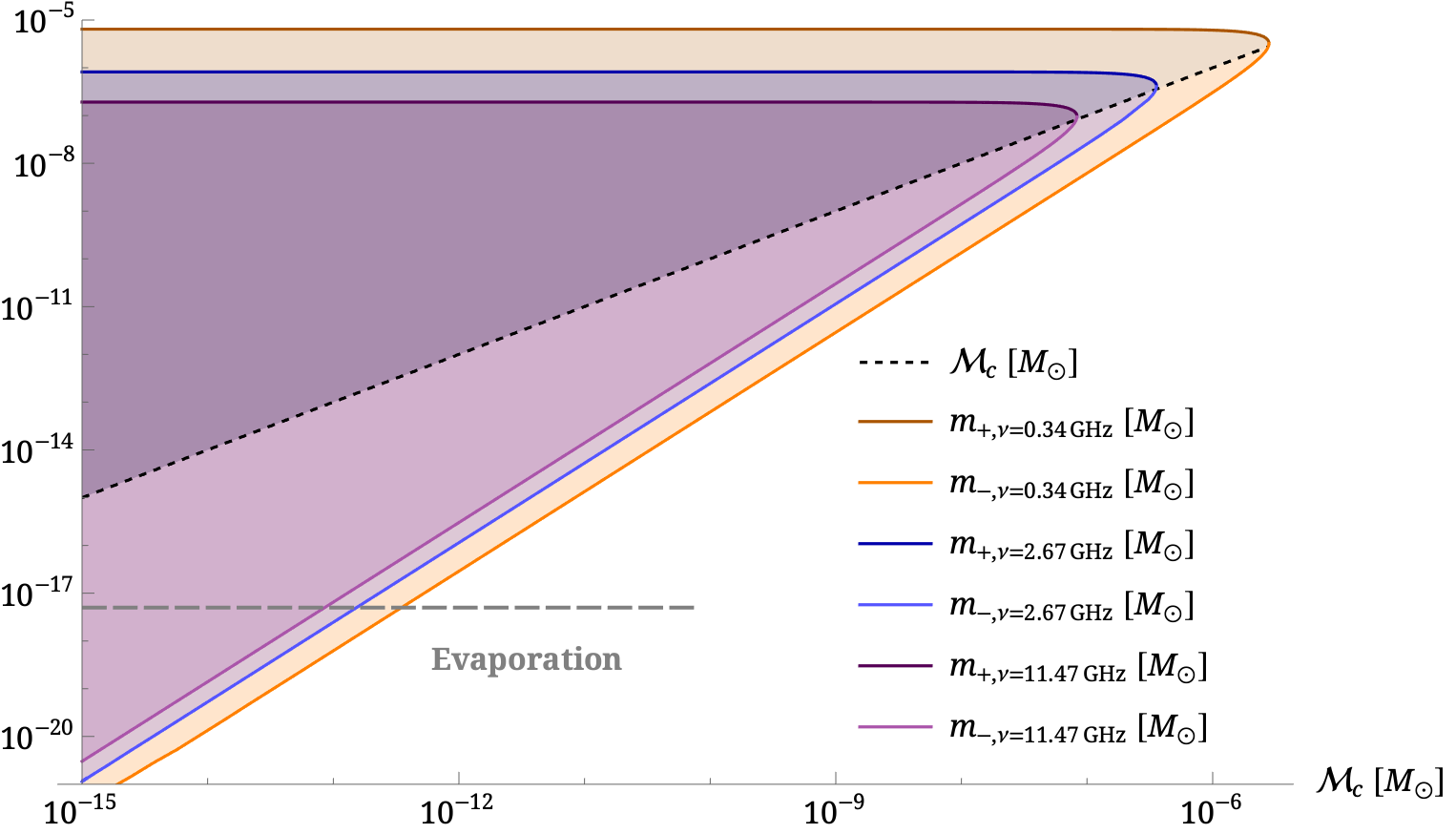}
    \caption{Minimum and maximum individual masses as a function of the chirp mass, under the constraint that the system has not yet reached its ISCO at the detection frequency.}%
    \label{fig2}
\end{figure}

This deserves a closer look.  Once more, it is worth emphasizing that, in principle, the situation is doubly degenerated. For a given detection frequency, there is a huge range of possible chirp masses and, in addition, for each chirp mass $\Mc$, there is a wide range of allowed individual masses $m_1$ and $m_2$. 
An obvious lower bound on the $m_i$'s is $M_*$ (which translates into an upper bound on the other mass). A more subtle upper bound is obtained by the following reasoning. As previously reminded, for a given $\Mc$, pulling the masses away from $m_1=m_2$ increases $M$ and the therefore the ISCO radius. This is acceptable unless the ISCO radius becomes so large that the associated frequency is now lower than the detection one. In this case the observable frequency is simply never reached in the evolution of the system. This sets an upper limit on the largest mass of the binary system. Figure \ref{fig2} displays this effect. For every chirp mass compatible with the considered experiment the maximum and minimum individual masses are evaluated. As expected, when the chirp mass approaches the maximum possible one, the system cannot handle a large asymmetry, it would otherwise already be inside its ISCO.

The extreme possible individual masses for a given chirp mass can be analytically calculated, at a fixed detection frequency $\nu$:

\begin{eqnarray}
m_{\pm}(\nu,\Mc) = \frac{c^3}{24\sqrt{6} \pi G \nu} \times \left( 1 \pm \sqrt{ 1-576 \sqrt{3} \times 2^{\frac{1}{6}} \left( \frac{\pi G  \nu \Mc}{c^3} \right)^{\frac{5}{3}}} \right).
\end{eqnarray}




This corresponds to the plain lines on Fig. \ref{fig2}. To summarize: the first figure of this article shows the wide range of chirp masses compatible with a detection configuration whereas the second figure exhibits the wide range of individual masses compatible with each of those chirp masses.\\

\section{Distances}

The post-Newtonian approximation for the strain induced by a source at distance $D$ and detected at frequency $\nu$ fixed by the experimental setup reads \cite{Maggiore:2007ulw}:

\begin{equation}
h=\frac{2}{D}\left( \frac{G\Mc}{c^2} \right)^{\frac{5}{3}} \left( \frac{\pi \nu}{c} \right)^{\frac{2}{3}}.
\label{hsource}
\end{equation}

As is well known, it exhibits a  linear dependence 
on the inverse distance\footnote{The energy scales, as usual, as $1/D^2$ but gravitational wave detectors do not ``absorb" the signal, hence they see the amplitude and the sensitivity to sources does scale, quite unusually, as $1/D$.}. In agreement with intuition, it also increases with the chirp mass and with the frequency, that is, for a given system, when approaching the merging phase. This strain produced by the source is to be compared with the minimum strain detectable by the considered experiment. Following \cite{Berlin:2021txa}, we write the signal power as

\be
\label{eq:Psig}
P_{\text{sig}} = \frac{1}{2 \mu_0 c^2} Q (2\pi \nu)^3 \Vcav^{\frac{5}{3}} (\eta h B_0)^2,
\ee
with $\mu_0$ the vacuum magnetic permeability, $\nu$ the frequency of the cavity mode $TM_{010}$, $B_0$ the magnetic field, $Q$ the cavity quality factor , $\Vcav$ the cavity volume and $\eta$ a coupling coefficient characterizing the interaction between the (dimensionless) effective current generated by a gravitational wave inside the cavity and a certain resonant mode, taken to be of order 0.1, as estimated in \cite{Berlin:2021txa}. In the case of very brief signals, this formula should however by modified by taking the signal quality factor instead of the cavity one. In the following the cavity quality factor is kept and this effect is accounted for through a modification of the effective time of the signal, as detailed later on.

\subsection{Conservative estimate}

In order to match with axionic setups and following \cite{Berlin:2021txa} we estimate the signal-to-noise ratio (SNR) thanks to the Dicke radiometer equation \cite{Sikivie:2020zpn, Berlin:2021txa, Berlin:2022hfx},

\begin{equation}
\label{eq:SNR1}
    {\rm SNR} \sim \frac{P_{sig}}{k_B T_{\text{sys}}} \,  \sqrt{\frac{\teff}{\Delta \nu}},
 \end{equation}
 where $\Delta \nu$ is the resolution bandwidth, $T_{\text{sys}}$ is the system temperature (including all contributions), $k_B$ the Boltzmann constant and $\teff$ is an effective time which depends on the physical situation considered, as will be made clear in the following. It coincides with the integration time only when dealing with systems observed long before their merging.





Obviously, Eq. (\ref{eq:SNR1}) is not the final word on the question of the SNR, and a room for improvement can be expected, especially when considering other detection methods \cite{singlePhotonAmplifierDetector}. We nevertheless make it our baseline for calculations so as to share a common background with other studies on this topic. As it will be made clear later, our conclusions do not depend on the detailed expression used for the SNR. We therefore rely on the Dicke equation and push it to its limit. This is not the only way to go but this is the working hypothesis of this article.\\

 Requiring SNR $>1$ leads\footnote{In practice, it might of course be necessary to increase the SNR beyond unity. This is straightforward to implement from our analysis and this would not change the conclusion.}, in the quasi-circular approximation, to
 
 \begin{equation}
\label{hdet}
 h>\sqrt{2 \mu_0 c^2 k_B}(2\pi \nu)^{-\frac{3}{2}} \eta^{-1} B_0^{-1} \Vcav^{-\frac{5}{6}} Q^{-\frac{1}{2}} T_{\text{sys}}^{\frac{1}{2}} \Delta \nu ^{\frac{1}{4}} \teff^{-\frac{1}{4}}.
 \end{equation}
 
 To fix orders of magnitude, we first assume that the spectral width entering this calculation is actually the bandwidth of the instrument, that is $\Delta \nu \sim \nu/Q$. We then consider the optimum spectral width when the signal is more coherent than the cavity.
 Normalizing to the three sets of GrAHal fiducial values used for Fig.\ref{fig1}, we obtain:


\begin{eqnarray}
\label{eq:sensitivity_estimate1}
h &>& 4.7 \times 10^{-22} \times \bigg( \frac{0.34 \ \text{GHz}}{\nu} \bigg)^{\frac{5}{4}} \bigg( \frac{0.1}{\eta} \bigg) \bigg( \frac{9 \ \text{T}}{B_0} \bigg) \bigg( \frac{5.01 \times 10^{-1} \ \text{m}^3}{\Vcav} \bigg)^{\frac{5}{6}} \bigg( \frac{10^5}{Q} \bigg)^{\frac{3}{4}} \bigg( \frac{T_\text{sys}}{0.3 \ \text{K}} \bigg)^{\frac{1}{2}} \bigg( \frac{1 \ \text{s}}{\teff} \bigg)^{\frac{1}{4}} \\ \nonumber
\Leftrightarrow h &>& 1.5 \times 10^{-21} \times \bigg( \frac{2.67 \ \text{GHz}}{\nu} \bigg)^{\frac{5}{4}} \bigg( \frac{0.1}{\eta} \bigg) \bigg( \frac{27 \ \text{T}}{B_0} \bigg) \bigg( \frac{1.83 \times 10^{-3} \ \text{m}^3}{\Vcav} \bigg)^{\frac{5}{6}} \bigg( \frac{10^5}{Q} \bigg)^{\frac{3}{4}} \bigg( \frac{T_\text{sys}}{0.4 \ \text{K}} \bigg)^{\frac{1}{2}} \bigg( \frac{1 \ \text{s}}{\teff} \bigg)^{\frac{1}{4}} \\ \nonumber
\Leftrightarrow h &>& 4.8 \times 10^{-21} \times \bigg( \frac{11.47 \ \text{GHz}}{\nu} \bigg)^{\frac{5}{4}} \bigg( \frac{0.1}{\eta} \bigg) \bigg( \frac{43 \ \text{T}}{B_0} \bigg) \bigg( \frac{4.93 \times 10^{-5} \ \text{m}^3}{\Vcav} \bigg)^{\frac{5}{6}} \bigg( \frac{10^5}{Q} \bigg)^{\frac{3}{4}} \bigg( \frac{T_\text{sys}}{1.0 \ \text{K}} \bigg)^{\frac{1}{2}} \bigg( \frac{1 \ \text{s}}{\teff} \bigg)^{\frac{1}{4}}~.
\end{eqnarray}

Of course, those formulas are exactly the same ones. We chose to write  times the very same thing so that the orders of magnitude can be immediately obtained for fiducial configurations and unit parameters.
Of course, those formulas are exactly the same ones. We chose to write three times the very same thing so that the orders of magnitude can be immediately obtained for fiducial configurations and unit parameters.
This allows a direct and simple comparison with the estimations performed in \cite{Berlin:2021txa} (taking into account that we normalized to $\teff = 1$ s and not $\teff = 1$ min). However,
the choice of the effective time $\teff$ is far from being straightforward and deserves a specific analysis as its expression depends upon the considered case and hides some crucial physical quantities. We will investigate this in details and will not leave it as a free parameter. This is one of the key-points of this study.\\

To this aim, we introduce the duration of the signal within the experimental bandwidth $\Delta \nu$, that is $t_{\Delta \nu} \sim \Delta \nu / \dot{f}(\nu)= \nu/(Q \dot{f}(\nu))$ where 
\begin{equation}
\label{fpoint}
\dot{f}(\nu)=\frac{96}{5}\pi^{\frac{8}{3}}\left( \frac{G\Mc}{c^3} \right)^{\frac{5}{3}}\nu^{\frac{11}{3}},
 \end{equation}
is the time derivative of the emitted gravitational wave frequency taken at the experimental frequency. This leads to
\begin{equation}
t_{\Delta \nu}\sim\frac{5}{96} \pi^{-\frac{8}{3}} \nu^{-\frac{8}{3}}Q^{-1}\left( \frac{G\Mc}{c^3} \right)^{-\frac{5}{3}}.
\label{t_int}
\end{equation}

\begin{figure}[!h]
    \centering
    \includegraphics[width=0.67\textwidth]{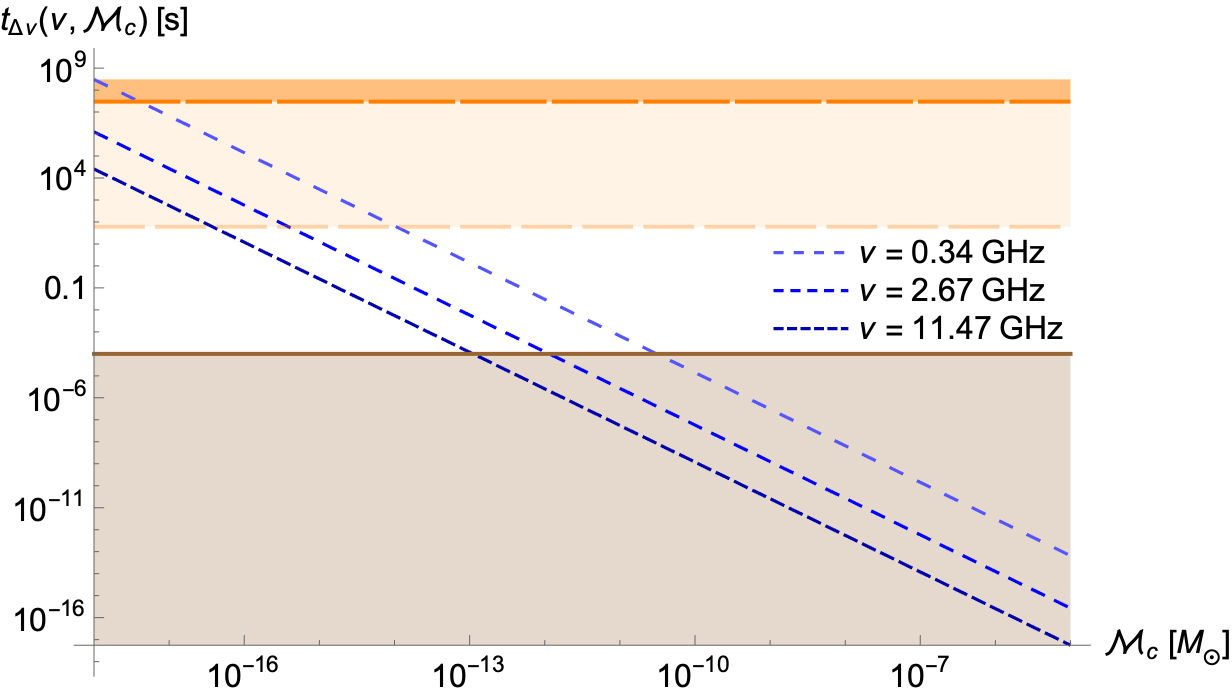}
    \caption{Time spent by the physical signal within the experimental bandwidth as a function of the chirp mass, for the three frequencies of interest, with a quality factor $Q=10^5$. \underline{Upper dashed horizontal line}: limit corresponding to an integration time of one year. \underline{Middle dashed horizontal line}: limit corresponding to an integration time of one minute. \underline{Lower plain horizontal line}: limit corresponding to $t_{\text{min}} \sim \Delta \nu ^{-1} \approx 10^{-4}$ s.}%
    \label{fig3}
\end{figure}

The value of $t_{\Delta \nu}$ as a function of the chirp mass is shown on Fig. \ref{fig3}. The upper ``deep" orange zone corresponds to $t_{\Delta \nu}>t_{\text{max}}\approx$ 1 year, that is longer that the full duration of the experiment whereas the upper ``light" orange zone corresponds to $t_{\Delta \nu}>t_{\text{max}}\approx$ 60 s. They are extreme cases for the maximum time (respectively for the full data taking and for a single run). The lower brown zone corresponds to a time spent by the astrophysical signal in the bandwidth smaller than the inverse of the sampling frequency of the experiment, that is $t_{\Delta \nu}<t_{\text{min}}\sim Q/\nu$, as it will be explained in more details later on. Clearly, the dependence of the time spent by the signal within the instrument bandwidth upon the chirp mass is not weak and cannot be ignored.\\

It is important to emphasize that a naive use of the formulas given, {\it e.g.}, in the reference article \cite{Berlin:2021txa} might lead to integrate during 1 minute -- taken as the fiducial time-scale for the experimental setup -- a signal whose real physical duration is less that a nanosecond! This is why the time analyses performed here is mandatory in deriving realistic estimates. The chirp mass is {\it not} independent of the possible integration time.\\

We now estimate in detail the minimum detectable strain and the maximum accessible distance for all physical cases. To keep the formalism rigorous and to make every step as clear as possible, we derive very general formulas and comment on the (obviously tiny) numerical values only in the discussion section.

\subsubsection{Effective time given by the signal frequency drift}

This case corresponds to $t_{\text{min}}<t_{\Delta \nu}<t_{\text{max}}$. It means that the effective time entering the calculation of the SNR is the duration of the signal within the frequency bandwidth of the cavity.
The associated range of masses is $\Mc \in \left[ \mathcal{M}^{\Delta\nu}_\text{min},\mathcal{M}^{\Delta\nu}_\text{max} \right]$, with:

\begin{eqnarray}
\mathcal{M}^{\Delta\nu}_\text{min} &=& \left( \frac{5}{96} \right)^{\frac{3}{5}} \pi^{-\frac{8}{5}} c^3 G^{-1} \nu^{-\frac{8}{5}} Q^{-\frac{3}{5}} t_{\text{max}}^{-\frac{3}{5}}  \\ \nonumber
&=& 3.9 \times 10^{-16} \times \left( \frac{2.67 ~ \text{GHz}}{\nu} \right)^{\frac{8}{5}} \left(\frac{10^5}{Q} \right)^{\frac{3}{5}} \left( \frac{60 ~ \text{s}}{t_{\text{max}}} \right)^{\frac{3}{5}} ~ M_\odot ~,
\end{eqnarray}

and 

\begin{eqnarray}
\label{mminmax}
\mathcal{M}^{\Delta\nu}_\text{max} &=& \ \left( \frac{5}{96} \right)^{\frac{3}{5}} \pi^{-\frac{8}{5}} c^3 G^{-1} \nu^{-\frac{8}{5}} Q^{-\frac{3}{5}} t_{\text{min}}^{-\frac{3}{5}}  \\ \nonumber
&=& 2.0 \times 10^{-12} \times \left( \frac{2.67 ~ \text{GHz}}{\nu} \right)^{\frac{8}{5}} \left(\frac{10^5}{Q} \right)^{\frac{3}{5}} \left( \frac{3.7 \times 10^{-5} ~ \text{s}}{t_{\text{min}}} \right)^{\frac{3}{5}} ~ M_\odot ~.
\end{eqnarray}

As it will be discussed in the last section of this study, one should notice that $t_{\text{min}}$ actually depends upon $Q$ and $\nu$.\\

In this situation, $\teff=t_{\Delta \nu}$ and Eq. (\ref{t_int}) can be used directly. Requiring $SNR>1$ leads to


\begin{eqnarray}
\label{hmin physical case}
h &>&  \left( \frac{6}{5} \right)^{\frac{1}{4}} \pi^{-\frac{5}{6}} \mu_0^{\frac{1}{2}} k_B^{\frac{1}{2}} c^{-\frac{1}{4}} G^{\frac{5}{12}}   \nu^{-\frac{7}{12}} \eta^{-1} B_0^{-1} \Vcav^{-\frac{5}{6}} Q^{-\frac{1}{2}} T_{\text{sys}}^{\frac{1}{2}} \Mc^{\frac{5}{12}}  \\ \nonumber 
\Leftrightarrow h &>& 2.0 \times 10^{-21} \times \left( \frac{2.67 ~ \text{GHz}}{\nu} \right)^{\frac{7}{12}} \left( \frac{0.1}{\eta} \right) \left( \frac{27 ~ \text{T}}{B_{0}} \right) \left( \frac{1.83 \times 10^{-3} \text{m}^3}{\Vcav} \right)^{\frac{5}{6}} \left(\frac{10^5}{Q} \right)^{\frac{1}{2}} \left( \frac{T_{\text{sys}}}{0.4 ~ \text{K}} \right)^{\frac{1}{2}}  \left( \frac{\Mc}{10^{-14} M_\odot} \right) ^{\frac{5}{12}}.
\end{eqnarray}

For the other setups considered here, the numerical prefactor becomes respectively: $1.6 \times 10^{-22}$ for the $0.34$ GHz configuration and $1.8 \times 10^{-20}$ for the $11.47$ GHz configuration (when taking care to normalize each time the magnetic field, the volume and the temperature to the corresponding proper fiducial values given in table \ref{Table GrAHal configurations}. This estimate (together with the ones given for the other cases in the following sections) is to be preferred over the proposal of Eq. (\ref{eq:sensitivity_estimate1}) where correlations between parameters are hiden, leading to misleading feelings about the functional dependencies.\\

By comparing this requirement with the the signal amplitude given by Eq. (\ref{hsource}), it is possible to derive the maximum distance at which the source can be situated to be detected : 

\begin{eqnarray}
\label{Dmax physical case}
D &<& \left( \frac{40}{3} \right)^{\frac{1}{4}} ~ \pi^{\frac{3}{2}} \mu_0^{-\frac{1}{2}} k_B^{-\frac{1}{2}} c^{-\frac{15}{4}} G^\frac{5}{4} \nu^{\frac{5}{4}} ~ \eta ~ B_0 ~ \Vcav^{\frac{5}{6}} Q^{\frac{1}{2}} ~ T_{\text{sys}}^{-\frac{1}{2}} ~ (\Mc)^{\frac{5}{4}} \\ \nonumber
\Leftrightarrow D &<& 8.1 \times 10^{3} \times \left( \frac{\nu}{2.67 ~ \text{GHz}} \right)^{\frac{5}{4}} \left( \frac{\eta}{0.1} \right) \left( \frac{B_{0}}{27 \ \text{T}} \right) \left( \frac{\Vcav}{1.83 \times 10^{-3} \ \text{m}^3} \right)^{\frac{5}{6}}  \left(\frac{Q}{10^5} \right)^{\frac{1}{2}} \times \left( \frac{0.4 ~ \text{K}}{T_{\text{sys}}} \right)^{\frac{1}{2}}  \left(\frac{\Mc}{10^{-14} M_\odot} \right) ^{\frac{5}{4}} ~ \text{m}
\end{eqnarray}

The prefactor for the other setups considered here are: $2.5 \times 10^{4} ~ \text{m}$ for the $0.34$ GHz configuration and $2.5 \times 10^{3} ~ \text{m}$ for the $11.47$ GHz configuration (still normalizing correctly the associated experimental quantities). 

Once again, we emphasize that this formula differs from what is sometimes encountered in the literature (not only because of prefactors but also in the exponents of the physical quantities). This is because it accounts for the fact that the time entering Eq. (\ref{eq:sensitivity_estimate}) cannot be set independently of the physical process considered. This is not a small correction but a huge effect. Importantly, the fact that lighter black holes are seen earlier in their in-spiralling history, and therefore spend more time within the detector bandwidth (as the frequency is evolving more slowly), does not counter-balance the decrease in signal intensity. As a result, the mass appears with a positive -- and even larger than 1 -- power in the distance evaluation.\\




\subsubsection{Effective time limited by the duration of the experiment}

This case corresponds to $t_{\Delta \nu}>t_{\text{max}}$. This means that the signal would spend ``more time than available" within the cavity bandwidth. This corresponds to very small chirp masses, such that $\mathcal{M}_C < \mathcal{M}^{\Delta\nu}_{\text{min}}$.



For such small masses, the signal is very stable. This is not intrinsic to the considered systems but due to the fact that, the frequency being fixed, the inspiralling process is in this case observed long before the merging. To evaluate the SNR, one must now set $\teff=t_{\text{max}}$. At this stage we however keep $\Delta \nu = \nu / Q$, with $Q$ the cavity quality factor. 

\begin{eqnarray}
\label{eq:sensitivity_estimate}
h &>& \frac{1}{2}\pi^{-\frac{3}{2}}\mu_0^{\frac{1}{2}} k_B^{\frac{1}{2}} c \nu^{-\frac{5}{4}}\eta^{-1}B_0^{-1}\Vcav^{-\frac{5}{6}}Q^{-\frac{3}{4}}T_{\text{sys}}^{\frac{1}{2}}t_{\text{max}}^{-\frac{1}{4}},\\ \nonumber
\Leftrightarrow h &>& 5.3 \times 10^{-22} \times \bigg( \frac{2.67 \ \text{GHz}}{\nu} \bigg)^{\frac{5}{4}} \bigg( \frac{0.1}{\eta} \bigg) \bigg( \frac{27 \ \text{T}}{B_0} \bigg) \bigg( \frac{1.83 \times 10^{-3} \ \text{m}^3}{\Vcav} \bigg)^{\frac{5}{6}} \bigg( \frac{10^5}{Q} \bigg)^{\frac{3}{4}} \bigg( \frac{T_\text{sys}}{0.4 \ \text{K}} \bigg)^{\frac{1}{2}} \bigg( \frac{60 \ \text{s}}{t_{\text{max}}} \bigg)^{\frac{1}{4}} ~,
\end{eqnarray}

with prefactors $1.7 \times 10^{-22}$ for the $0.34$ GHz configuration and $1.7 \times 10^{-21}$ for the $11.47$ GHz configuration. Importantly, the minimum required strain does {\it not} depend on the chirp mass in this case. This is due to fact that the effective duration does indeed not depend upon $\Mc$.

This leads to:

\begin{eqnarray}
D &<&  4 \pi^{\frac{13}{6}} \mu_0^{-\frac{1}{2}} k_B^{-\frac{1}{2}} c^{-5} G^{\frac{5}{3}} \nu^{\frac{23}{12}} \eta B_0 \Vcav^{\frac{5}{6}} Q^{\frac{3}{4}} T_{\text{sys}}^{-\frac{1}{2}} t_{\text{max}}^{\frac{1}{4}} \Mc^{\frac{5}{3}}~. \\ \nonumber
\Leftrightarrow D &<& 0.31
\times \left( \frac{\nu}{2.67 ~ \text{GHz}} \right)^{\frac{23}{12}} \left( \frac{\eta}{0.1} \right) \left( \frac{B_{0}}{27 \ \text{T}} \right) \left( \frac{\Vcav}{1.83 \times 10^{-3} \ \text{m}^3} \right)^{\frac{5}{6}}  \left(\frac{Q}{10^5} \right)^{\frac{3}{4}} \left( \frac{0.4 ~\text{K}}{T_{\text{sys}}} \right)^{\frac{1}{2}} \left( \frac{t_{\text{max}}}{60 ~ \text{s}} \right)^{\frac{1}{4}} \left( \frac{\Mc}{10^{-17} M_\odot} \right) ^{\frac{5}{3}} ~ \text{m}.
\end{eqnarray}

The prefactors for the other setups are: $0.25 ~ \text{m}$ for the $0.34$ GHz configuration and also $0.25 ~ \text{m}$ for the $11.47$ GHz configuration. 
For a haloscope experiment to observe coalescences of so light black holes, the merger would have to take place inside the resonant cavity. Beside the fact that the reasoning applied to compute the signal would clearly break down, this makes no sense. This will be shown in more details in the following sections. Interestingly, from a purely formal point of vue, the two extreme configurations lead to the same accessible distance and the 2.67 GHz is the one corresponding to the largest probed volume.\\

It should be noticed that the chirp mass now enters the maximum distance with an even higher power than in the previous case. This does not come as a surprise as the time integration effect that was playing in the favor of small masses does not play any role in this case.\\ 

Obviously, this make the situation even worse than in the previous case, as it is not even possible to take advantage of the full duration of the signal. If one sets $t_{\text{max}}\approx 1$ yr or even the time corresponding to black holes with $m=M_*$, the results will be catastrophic in the entire range covered by this case. (In practice 1 yr is probably not realistic as, for a very long duration, an additional pink noise should be taken into account.) If, however, the maximum integration time is, for experimental reasons, of the order of one minute, this new formula can be used to estimate accurate numbers that might, {\it a priori} not be fully out of reach. 

\subsubsection{Effective time limited by $t_{\Delta \nu}$}

This is the most interesting regime as it corresponds to chirp masses $\mathcal{M}_C > \mathcal{M}^{\Delta\nu}_{\text{max}}$, that is to a vast portion of the parameter space which includes the highest masses accessible. To properly estimate the sensitivity, different limiting criteria must be taken into account, as the Dicke approach is pushed to its limit. Those effects are discussed below and added one by one to better illustrate how they decrease the sensitivity at high masses.\\

i) \underline{Physics limited case}\\

We start by ignoring the instrumental effects and we consider an ideal (and unrealistic) case where the limit is only set by the physical drift in time of the signal frequency. We recall that even this obvious effect was sometimes neglected. This is exactly the same situation as in the previous case and the very same formulas, Eqs. (\ref{hmin physical case}) and (\ref{Dmax physical case}), can be used. We simply allow them to be pushed to larger masses, namely to chirp masses $\Mc > \mathcal{M}_{\text{max}}^{\Delta \nu} \approx 10^{-12} M_\odot $. We however emphasize that the chirp mass can, anyway, not be too high, otherwise the black holes would be located inside their ISCO (or would even already have merged) before emitting gravitational waves in the GHz range, as shown in Fig. \ref{fig1}.\\

ii) \underline{Short $t_{\Delta \nu}$ effect}\\

For chirp masses such that $\mathcal{M}_C > \mathcal{M}^{\Delta\nu}_{\text{max}}$ the time spent by the signal in the bandwidth of the detector $t_{\Delta \nu}$, given as usual by Eq. (\ref{t_int}), becomes shorter than $t_{\text{min}} \approx \Delta \nu^{-1}$. The signal in the cavity lasts for $t_{\text{min}}$ (cavity decay time), hence $t_{\text{eff}} = t_{\text{min}}$. However the signal is only received during a fraction $\teff=t_{\Delta \nu}/t_{\text{min}}$ of the full measurement time. Compared to the case of a stationary signal, the mean power is thus reduced by the same factor. The effective time entering the Dicke equation then becomes $\teff=t_{\Delta \nu}^2/t_{\text{min}}$. 



Taking this effect into account and requiring $SNR>1$ leads to a minimum strain for detection:


\begin{eqnarray}
h &>& \left( \frac{24}{5} \right)^{\frac{1}{2}} \pi^{-\frac{1}{6}} c^{-\frac{3}{2}} G^{\frac{5}{6}} \mu_0^{\frac{1}{2}} k_B^{\frac{1}{2}} \nu^{-\frac{1}{6}} \eta^{-1} B_0^{-1} \Vcav^{-\frac{5}{6}}  T_{\text{sys}}^{\frac{1}{2}} \Mc^{\frac{5}{6}} \\ \nonumber 
\Leftrightarrow h &>& 3.3 \times 10^{-18} \times \bigg( \frac{2.67 \ \text{GHz}}{\nu} \bigg)^{\frac{1}{6}} \bigg( \frac{0.1}{\eta} \bigg) \bigg( \frac{27 \ \text{T}}{B_0} \bigg) \bigg( \frac{1.83 \times 10^{-3} \ \text{m}^3}{\Vcav} \bigg)^{\frac{5}{6}} \bigg( \frac{T_\text{sys}}{0.4 \ \text{K}} \bigg)^{\frac{1}{2}} \left( \frac{\Mc}{10^{-9} M_\odot} \right) ^{\frac{5}{6}}~.
\end{eqnarray}

The values of the prefactors for the other setups are: $1.1 \times 10^{-19}$ for the $0.34$ GHz configuration and $5.2 \times 10^{-17}$ for the $11.47$ GHz configuration.\\


In this case the reachable distance is:


\begin{eqnarray}
D &<& \sqrt{\frac{5}{6}} \pi^{\frac{5}{6}} c^{-\frac{5}{2}} G^{\frac{5}{6}}  \mu_0^{-\frac{1}{2}} k_B^{-\frac{1}{2}} \nu^{\frac{5}{6}} \eta B_0 V_{\text{cav}}^{\frac{5}{6}} T_{\text{sys}}^{-\frac{1}{2}} \Mc^{\frac{5}{6}}\\ \nonumber
\Leftrightarrow D &<& 1.1 \times 10^{9} \times \left( \frac{\nu}{2.67 ~ \text{GHz}} \right)^{\frac{5}{6}} \left( \frac{\eta}{0.1} \right) \left( \frac{B_{0}}{27 \ \text{T}} \right) \left( \frac{\Vcav}{1.83 \times 10^{-3} \ \text{m}^3} \right)^{\frac{5}{6}}   \left( \frac{0.4 ~\text{K}}{T_{\text{sys}}} \right)^{\frac{1}{2}} \left( \frac{\Mc}{10^{-9} M_\odot} \right) ^{\frac{5}{6}} ~ \text{m}.
\end{eqnarray}

The values for the other setups are: $8.1 \times 10^{9} ~ \text{m}$ for the $0.34$ GHz configuration and $1.8 \times 10^{8} ~ \text{m}$ for the $11.47$ GHz configuration.\\

iii) \underline{Cavity charging effect}\\

We now forget the previous effect and we focus on another effect limiting the sensitivity for short signals: the fact that the $Q$ factor entering the expression of the power, Eq. (\ref{eq:Psig}), should not be the cavity quality factor but the signal quality factor when the latter is smaller.
For masses such that $t_{\Delta \nu} < \Delta \nu^{-1}$, one therefore has to replace: $Q \rightarrow \nu t_{\Delta \nu} = Q t_{\Delta \nu} / t_{\text{min}}$ \cite{Kim:2020kfo}. This can be taken into account by setting $t_{eff}=t_{\Delta \nu}^3 / t_{\text{min}}^2$, which leads to:

\begin{eqnarray}
h &>& \frac{1}{2} \left( \frac{96}{5} \right)^{\frac{3}{4}} \pi^{\frac{1}{2}} \mu_0^{\frac{1}{2}} k_B^{\frac{1}{2}} c^{-\frac{11}{4}} G^{\frac{5}{4}}   \nu^{\frac{1}{4}} \eta^{-1} B_0^{-1} \Vcav^{-\frac{5}{6}} Q^{\frac{1}{2}} T_{\text{sys}}^{\frac{1}{2}} \Mc^{\frac{5}{4}}  \\ \nonumber 
\Leftrightarrow h &>& 4.3 \times 10^{-17} \times \left( \frac{\nu}{2.67 ~ \text{GHz}} \right)^{\frac{1}{4}} \left( \frac{0.1}{\eta} \right) \left( \frac{27 ~ \text{T}}{B_{0}} \right) \left( \frac{1.83 \times 10^{-3} \text{m}^3}{\Vcav} \right)^{\frac{5}{6}} \left(\frac{Q}{10^5} \right)^{\frac{1}{2}} \left( \frac{T_{\text{sys}}}{0.4 ~ \text{K}} \right)^{\frac{1}{2}}  \left( \frac{\Mc}{10^{-9} M_\odot} \right) ^{\frac{5}{4}}.
\end{eqnarray}

The numerical prefactors for the other setups are: $6.2 \times 10^{-19}$ for the $0.34$ GHz configuration and $1.2 \times 10^{-15}$ for the $11.47$ GHz configuration.\\

The associated maximal distance at which a merger could be observed is thus:

\begin{eqnarray}
D &<& 4 \bigg( \frac{5}{96} \bigg)^{\frac{3}{4}} \pi^{\frac{1}{6}} c^{-\frac{5}{4}} G^{\frac{5}{12}}  \mu_0^{-\frac{1}{2}} k_B^{-\frac{1}{2}} \nu^{\frac{5}{12}} \eta B_0 V_{\text{cav}}^{\frac{5}{6}} Q^{-\frac{1}{2}} T_{\text{sys}}^{-\frac{1}{2}} \Mc^{\frac{5}{12}} \\ \nonumber
\Leftrightarrow D &<& 8.3 \times 10^{7} \times \left( \frac{\nu}{2.67 ~ \text{GHz}} \right)^{\frac{5}{12}} \left( \frac{\eta}{0.1} \right) \left( \frac{B_{0}}{27 \ \text{T}} \right) \left( \frac{\Vcav}{1.83 \times 10^{-3} \ \text{m}^3} \right)^{\frac{5}{6}} \left(\frac{10^5}{Q} \right)^{\frac{1}{2}}  \left( \frac{0.4 ~\text{K}}{T_{\text{sys}}} \right)^{\frac{1}{2}} \left( \frac{\Mc}{10^{-9} M_\odot} \right) ^{\frac{5}{12}} ~ \text{m}.
\end{eqnarray}

The numerical prefactors for the other setups are: $1.5 \times 10^{9}~ \text{m}$ for the $0.34$ GHz configuration and $7.5 \times 10^{6}~ \text{m}$ for the $11.47$ GHz configuration.\\

iv) \underline{Full effect}\\

To get a conservative sensitivity estimate for the "high masses range" we now take into account both previous effects simultaneously. This can be accounted for by setting $\teff=t_{\Delta \nu}^4/t_{\text{min}}^3$ (see the appendix for more details and comments on this treatment). 
Still requiring $SNR>1$ for detection, this  leads to a minimum strain for detection of:\\

\begin{eqnarray}
h &>& \frac{48}{5}  \pi^{\frac{7}{6}} c^{-4} G^{\frac{5}{3}} \mu_0^{\frac{1}{2}} k_B^{\frac{1}{2}} \nu^{\frac{2}{3}} \eta^{-1} B_0^{-1} \Vcav^{-\frac{5}{6}} Q  T_{\text{sys}}^{\frac{1}{2}} \Mc^{\frac{5}{3}} \\ \nonumber 
\Leftrightarrow h &>& 5.6 \times 10^{-16} \times \bigg( \frac{\nu}{2.67 \ \text{GHz}} \bigg)^{\frac{2}{3}} \bigg( \frac{0.1}{\eta} \bigg) \bigg( \frac{27 \ \text{T}}{B_0} \bigg) \bigg( \frac{1.83 \times 10^{-3} \ \text{m}^3}{\Vcav} \bigg)^{\frac{5}{6}} \bigg( \frac{Q}{10^5} \bigg) \bigg( \frac{T_\text{sys}}{0.4 \ \text{K}} \bigg)^{\frac{1}{2}} \left( \frac{\Mc}{10^{-9} M_\odot} \right) ^{\frac{5}{3}} ~.
\end{eqnarray}

The values of the numerical prefactors for the other setups are respectively: $3.4 \times 10^{-18}$ for the $0.34$ GHz configuration and $3.0 \times 10^{-14}$ for the $11.47$ GHz configuration.\\


The associated reachable distance becomes:


\begin{eqnarray}
D &<& \frac{5}{24} \pi^{-\frac{1}{2}} \mu_0^{-\frac{1}{2}} k_B^{-\frac{1}{2}} \eta B_0 V_{\text{cav}}^{\frac{5}{6}} Q^{-1} T_{\text{sys}}^{-\frac{1}{2}}\\ \nonumber
\Leftrightarrow D &<& 6.3 \times 10^{6} \times  \left( \frac{\eta}{0.1} \right) \left( \frac{B_{0}}{27 \ \text{T}} \right) \left( \frac{\Vcav}{1.83 \times 10^{-3} \ \text{m}^3} \right)^{\frac{5}{6}} \left(\frac{10^5}{Q} \right)  \left( \frac{0.4 ~\text{K}}{T_{\text{sys}}} \right)^{\frac{1}{2}}  ~ \text{m}.
\end{eqnarray}

The values for the other setups are: $2.6 \times 10^{8} ~ \text{m}$ for the $0.34$ GHz configuration and $3.1 \times 10^{5} ~ \text{m}$ for the $11.47$ GHz configuration.

\subsection{Optimized resolution bandwidths}

In cases where the signal is more coherent than the cavity ($\frac{1}{\Delta \nu} < t_{\Delta \nu} \Leftrightarrow  \dot{f} t_{\text{int}} < \frac{\nu}{Q}$), phase sensitive detectors allow to optimize the signal-to-noise ratio by setting the resolution bandwidth at the signal spectral width.\\

i) \underline{Effective time limited by the duration of the experiment}\\

In the "low-mass" regime corresponding to $t_{\Delta\nu} > t_{\text{max}}$ the signal spectral width is given by $\Delta\nu_{\text{signal}} = t_{\text{max}}^{-1}$ and the signal to noise ratio now reads:

\begin{equation}
    {\rm SNR}\sim \frac{P_{sig}}{k_BT_{\text{sys}}}t_{\text{int}}.
\end{equation}
This leads to 
 \begin{eqnarray}
 h &>& \frac{1}{2} \pi^{-\frac{3}{2}} c \mu_0^{\frac{1}{2}} k_B^{\frac{1}{2}} \nu^{-\frac{3}{2}} \eta^{-1} B_0^{-1} \Vcav^{-\frac{5}{6}} Q^{-\frac{1}{2}} T_{\text{sys}}^{\frac{1}{2}} t_{\text{max}}^{-\frac{1}{2}} \\ \nonumber
\Leftrightarrow h &>& 1.5 \times 10^{-23} \times \bigg( \frac{2.67 \ \text{GHz}}{\nu} \bigg)^{\frac{3}{2}} \bigg( \frac{0.1}{\eta} \bigg) \bigg( \frac{27 \ \text{T}}{B_0} \bigg) \bigg( \frac{1.83 \times 10^{-3} \ \text{m}^3}{\Vcav} \bigg)^{\frac{5}{6}} \bigg( \frac{10^5}{Q} \bigg)^{\frac{1}{2}} \bigg( \frac{T_\text{sys}}{0.4 \ \text{K}} \bigg)^{\frac{1}{2}} \bigg( \frac{60 \ \text{s}}{t_{\text{max}}} \bigg)^{\frac{1}{2}} ~,
 \end{eqnarray}
 which translates into
\begin{eqnarray}
D &<& 4 \pi^{\frac{13}{6}} c^{-5} G^{\frac{5}{3}}  \mu_0^{-\frac{1}{2}} k_B^{-\frac{1}{2}} \nu^{\frac{13}{6}} \eta B_0 V_{\text{cav}}^{\frac{5}{6}} Q^{\frac{1}{2}} T_{\text{sys}}^{-\frac{1}{2}} \Mc^{\frac{5}{3}} t_{\text{max}}^{\frac{1}{2}} \\ \nonumber
\Leftrightarrow D &<& 11 \times \bigg( \frac{\nu}{2.67 \ \text{GHz}} \bigg)^{\frac{13}{6}}  \left( \frac{\eta}{0.1} \right) \left( \frac{B_{0}}{27 \ \text{T}} \right) \left( \frac{\Vcav}{1.83 \times 10^{-3} \ \text{m}^3} \right)^{\frac{5}{6}} \left(\frac{Q}{10^5} \right)^{\frac{1}{2}}  \left( \frac{0.4 ~\text{K}}{T_{\text{sys}}} \right)^{\frac{1}{2}} \left( \frac{\Mc}{10^{-17} M_\odot} \right) ^{\frac{5}{3}} \bigg( \frac{t_{\text{max}}}{60 \ \text{s}} \bigg)^{\frac{1}{2}} ~ \text{m}.
\end{eqnarray}

ii) \underline{Effective time limited by the duration of the signal}\\

In the "intermediate-mass" regime for which $t_{\text{min}} < t_{\Delta\nu} < t_{\text{max}}$ the signal spectral width is given by $\Delta\nu_{\text{signal}} = t_{\Delta\nu}^{-1}$ and the signal to noise ratio now reads:

\begin{equation}
    {\rm SNR}\sim \frac{P_{sig}}{k_BT_{\text{sys}}}t_{\Delta\nu}.
\end{equation}
This leads to 
 \begin{eqnarray}
 h &>& \sqrt{\frac{24}{5}} \pi^{-\frac{1}{6}} c^{-\frac{3}{2}} G^{\frac{5}{6}} \mu_0^{\frac{1}{2}} k_B^{\frac{1}{2}} \nu^{-\frac{1}{6}} \eta^{-1} B_0^{-1} \Vcav^{-\frac{5}{6}} T_{\text{sys}}^{\frac{1}{2}} \Mc^{\frac{5}{6}} \\ \nonumber \Leftrightarrow h &>& 2.2 \times 10^{-22} \times \bigg( \frac{2.67 \ \text{GHz}}{\nu} \bigg)^{\frac{1}{6}} \bigg( \frac{0.1}{\eta} \bigg) \bigg( \frac{27 \ \text{T}}{B_0} \bigg) \bigg( \frac{1.83 \times 10^{-3} \ \text{m}^3}{\Vcav} \bigg)^{\frac{5}{6}} \bigg( \frac{T_\text{sys}}{0.4 \ \text{K}} \bigg)^{\frac{1}{2}} \left( \frac{\Mc}{10^{-14} M_\odot} \right) ^{\frac{5}{6}} ~,
 \end{eqnarray}
 which translates into
\begin{eqnarray}
D &<& \sqrt{\frac{5}{6}} \pi^{\frac{5}{6}} c^\frac{-5}{2} G^{\frac{5}{6}} \mu_0^{-\frac{1}{2}} k_B^{-\frac{1}{2}} \nu^{\frac{5}{6}} \eta B_0 V_{\text{cav}}^{\frac{5}{6}} T_{\text{sys}}^{-\frac{1}{2}} \Mc^{\frac{5}{6}} \\ \nonumber
\Leftrightarrow D &<& 7.4 \times 10^4 \times \bigg( \frac{\nu}{2.67 \ \text{GHz}} \bigg)^{\frac{5}{6}}  \left( \frac{\eta}{0.1} \right) \left( \frac{B_{0}}{27 \ \text{T}} \right) \left( \frac{\Vcav}{1.83 \times 10^{-3} \ \text{m}^3} \right)^{\frac{5}{6}}  \left( \frac{0.4 ~\text{K}}{T_{\text{sys}}} \right)^{\frac{1}{2}} \left( \frac{\Mc}{10^{-14} M_\odot} \right) ^{\frac{5}{6}} ~ \text{m}..
\end{eqnarray}

\section{Results and discussion}

\begin{figure}[!h]
\includegraphics[width=0.67\textwidth]{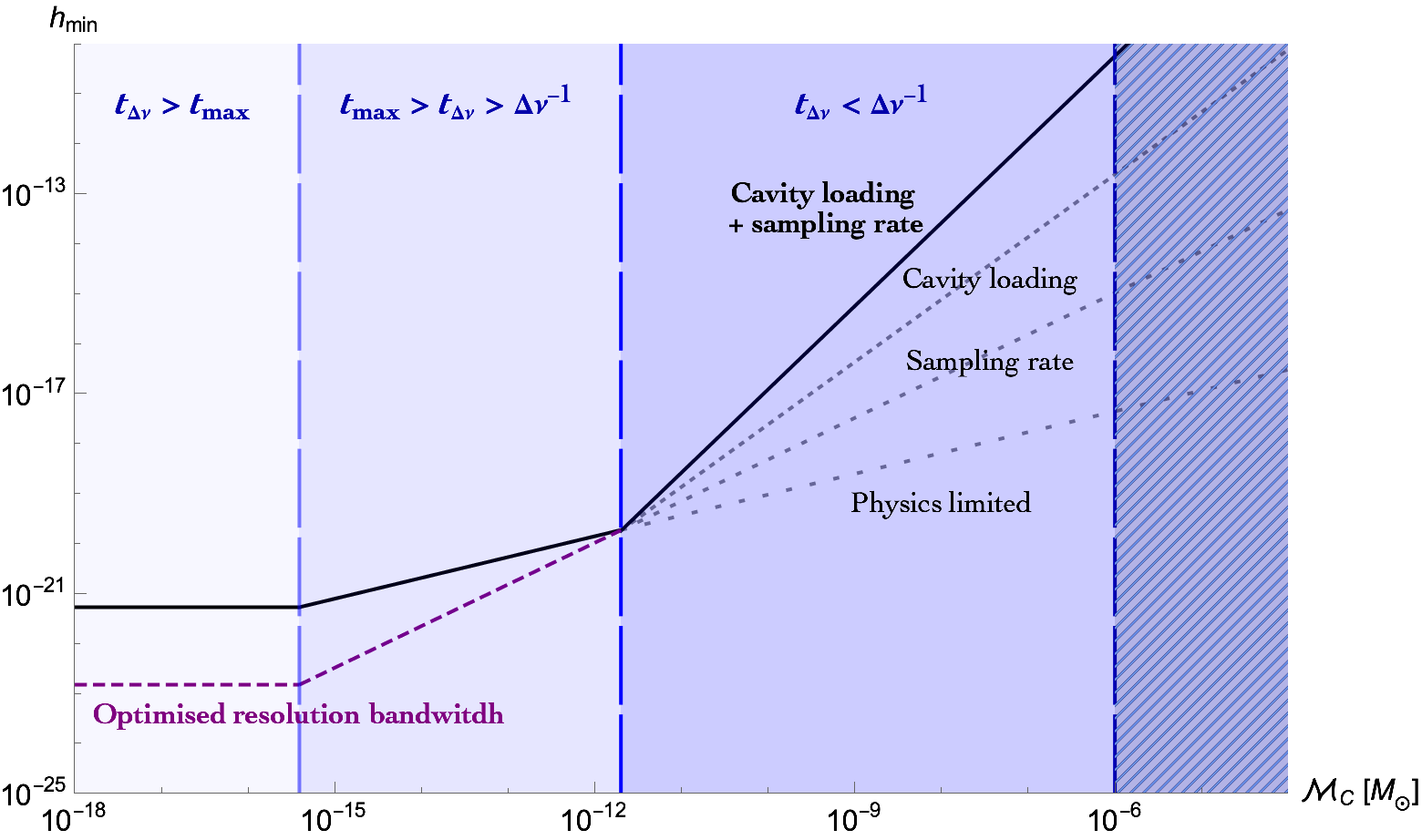}
\caption{Strain sensitivity of the GrAHal 2.67 GHz configuration as a function of the chirp mass of the physical system. The three different temporal regimes can be easily identified. The region above $10^{-6}$ solar mass is forbidden. The dashed line on the left part corresponds to an improvement relying on an optimized resolution bandwidth. The three dashed lines on the right part correspond to the different temporal effects -- accounted for individually -- described in the text. The dark plain line corresponds to the best possible estimate taking everything into account.
The values $\Delta \nu^{-1} = 10^{-4}$ s and $t_{\text{max}} = 60$ s have been chosen.}\label{figResultSensitivity}
\end{figure}

Figure \ref{figResultSensitivity} shows the strain sensitivity as a function of the chirp mass in all the considered regimes. The plain black line corresponds to the most ``realistic" case whereas dashed lines correspond to the different hypotheses discussed previously. It should be first emphasized once more that the strain required for the considered binary system to be detected {\it does} depend upon the (chirp) mass. This is because the time structure of the signal -- hence its interplay with the different timescales of the experiment -- highly depends on the mass. It is therefore  not sufficient to characterize a cavity experiment by a (minimum) strain to which it is sensitive. As we show here, for masses between $10^{-18}$ and $10^{-6}$ solar mass, the strain required for the signal to be detectable varies by 10 orders of magnitude. The higher the mass, the higher the required strain (as the signal becomes shorter). Impressively, for the highest detectable masses, corresponding to a system seen very close to merging, the required strain for the measurement to be possible is {\it very} high, around $h\sim 10^{-11}$, and {\it very} different from the sometime advertised sensitivity. The situation is obviously not the same for broadband experiment like Virgo and LIGO (see, e.g., Section 7 of \cite{Maggiore:2007ulw}).\\

\begin{figure}[!h]
\includegraphics[width=0.67\textwidth]{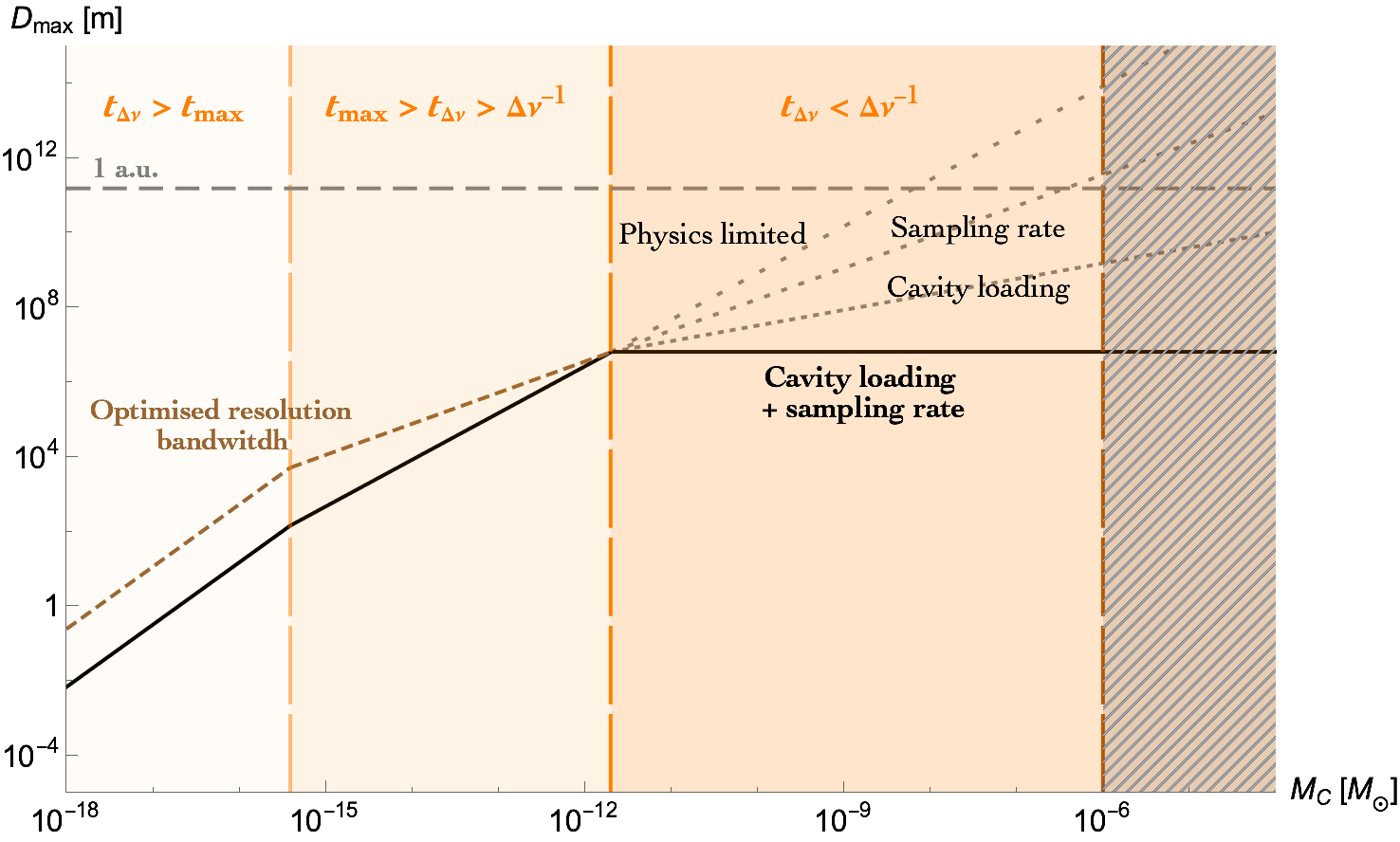}
\caption{Maximum distance to the inspiralling system for the GrAHal 2.67 GHz configuration as a function of the chirp mass of the physical system. The three different temporal regimes can easily be identified. The region above $10^{-6}$ solar mass is forbidden. The dashed line on the left part corresponds to an improvement relying to an optimized resolution bandwidth. The three dashed lines on the right part correspond to the different temporal effects -- accounted for individually -- described in the text. The dark plain line corresponds to the best possible estimate taking everything into account.
The values $\Delta \nu^{-1} = 10^{-4}$ s and $t_{\text{max}} = 60$ s have been chosen.}\label{figResultDistance}
\end{figure}

Figure \ref{figResultDistance} displays the maximum distance at which the binary system can de detected as a function of the chirp mass. The plain line still corresponds to the ``most" realistic case. Although the required strain is higher for larger masses, the distance remains an increasing function of the mass because the fast varying amplitude of signal plays a bigger role than the temporal effects, shown on the previous figure, which favors small masses. This was not {\it a priori} obvious. Interestingly, one can however notice that the accessible distance remains constant above a few times $10^{-12}$ solar mass. This is because, in this regime, the subtle experimental effects happen to exactly compensate for the physical increase in generated strain. We have previously argued that, should the sensitivity be smaller for smaller masses, it would still be mandatory to consider them as we do not know what are the actual masses of hypothetical light black holes around us. But this figure gives another argument: the sensitivity (in distance) happens to be flat between $2\times10^{-12}$ and $10^{-6}$ solar masses.

The most important and obvious result from this study is clear: beyond interesting ``trends" revealing nice physical effects, the numerical value of the maximum distance is {\it tiny}. At best, it reaches 1000 km. We are forced to conclude that no detection can be expected. It is not even worth calculating an expected event rate or trying to identify which parameters should be improved.\\

If one however plays the game to follow \cite{Clesse:2016ajp} and consider primordial black holes clustered in virialized sub-halos in a model fully characterized by the local density contrast $\delta$ (that can be taken close to $10^9$ for a rough order of magnitude), one is led to an event rate $\tau\sim 10^{-56}~ \text{yr}^{-1}$. Even if the formula we have used in this work for the SNR is deeply sub-optimal it seems hard to imagine that such a tiny number can be compensated for by any refined analysis.\\

\section{Other sources}

Beyond black holes and ignoring the extremely small numerical values obtained previously, quite a lot of compact objects could, in principle, be considered. Among other sources, one can think to conventional compact objects, boson stars, fermion stars, dark matter stars, multi-components stars, or dark energy stars \cite{Giudice:2016zpa}. 
We simply discuss here briefly the consequences of introducing a compactness parameter, in the sense of \cite{Alho:2022bki}, $C=Gm/(Rc^2)$ -- assumed to be the same for the two merging bodies, hence the mass $m$ and size $R$ -- for purely theoretical purposes, in case this study is used in another context.  \\

For black holes, the compactness is $C=1/2$. It can only be smaller for other sources. This obviously makes the situation even worse. In general, the radius of the innermost stable circular orbit is 
\begin{equation}
R_{\rm ISCO}=\frac{3GM}{Cc^2},    
\end{equation}
where, as before, $m$ is the total mass of the considered binary system. 
It is quite usual in the literature (see, {\it e.g.} \cite{Aggarwal:2020olq}) to express the strain for such a system (assuming that the two objects have the same mass $m$) as
\begin{equation}
    h\propto C\frac{m}{D}.
    \label{compact}
\end{equation}
It seems to be proportional to the compactness $C$. However, in a sense, this way of writing the strain is over-simplistic. The C-factor actually enters this expression of the strain only because (at leading order) it is implicitly written at the ISCO, which is obviously reached earlier for a less compact system. In fact, Eq. (\ref{compact}) gives the maximum strain for a given inspiralling system. This is not the relevant quantity for the analysis presented here as, as explained in the previous sections, the narrow bandwidth of the experiments in the considered frequency range makes it highly unlikely that the system is observed at the most favorable stage of its evolution and as, in addition, the full sensitivity does not even make this situation easier to detect.\\

As a consequence, we suggest instead to still consider the results of the previous sections as valid (that is with {\it no } appearance of the compactness at all in the formulas, even for objects with arbitrary $C-$factors), but to limit the parameter space to the range where the system has not reached its ISCO. Kepler's law reads
\begin{equation}
    \omega_s^2=G\frac{M}{R^3}.
\end{equation}
Requiring $R>R_{\rm ISCO}$, one immediately sees that the maximal mass allowed scales as $C^{3/2}$, namely (remembering that, due to their quadrupolar nature, the frequency of gravitational waves is twice the orbital frequency),
\begin{equation}
    M<\frac{c^3}{\pi \sqrt{27}Gf}C^{\frac{3}{2}}.
\end{equation}
Intuitively, at a given frequency, considering smaller masses requires to have the objects closer one to the other. But, as obvious from Kepler's law, the distance scales as the power 1/3 of the mass. On the other hand, the ISCO radius scales linearly with the mass. This is why the compactness is an issue for large masses. This basically means that the parameter space previously given simply has to be truncated so that
\begin{equation}
    m<\left( \frac{C}{1/2} \right)^{3/2}m_{\text{max}},
\end{equation}

where $m_{\text{max}}$ is the maximum allowed individual mass previously estimated. Within this limit (which is stringent for non-compact sources), the previous strain calculation, and therefore the distance estimates, still hold without correcting for the $C-$factor.
One should keep in mind that for most non-exotic objects the ISCO is located inside the considered object.

\section{Conclusion}

The fact that haloscopes, designed to search for axions, might allow the detection of ultra high frequency gravitational waves is quite amazing. This opens an exciting new window on the Universe. This is obviously a rapidly developing field with appealing features to search for new physics which might lead to surprises in astrophysics, cosmology, or particle physics.\\

There might exist interesting stochastic signals, like backgrounds from the early Universe (see references in \cite{Aggarwal:2020olq}), detectable by microwave cavities \cite{Herman:2022fau}. Signals from the late Universe can also be expected \cite{Aggarwal:2020olq}. Among others \cite{Arvanitaki:2014wva, Brito:2015oca, Tootle:2022pvd, Casalderrey-Solana:2022rrn}, it could also have been hoped that close hyperbolic encounters \cite{Garcia-Bellido:2017qal,Garcia-Bellido:2017knh,Morras:2021atg} lead to measurable gravitational waves in the frequency band considered here. This seems not to be the case \cite{Teuscher:2024xft,Barrau:2024kcb}.
Several other exotic ideas also deserve consideration. We leave their detailed investigation (among many other possible sources) for future works. 

As far as coherent signals from binary systems -- the topic of this work -- are concerned it however seems that any detection is very unlikely. \\

We have performed an in-depth estimate of the sensitivity in all physical regimes and underlined strong correlations between parameters that are sometimes inappropriately considered as independent and have shown how to take them into account.  In particular, temporal aspects have been carefully accounted for. We have calculated distances that can be probed and ours results exhibit that current experiments remain very far from the relevant sensitivities. This is our main conclusion: at this stage, no signal at all is to be expected from binary systems of black holes.\\

It would certainly be welcome to investigate in more details the signal-to-noise ratio which has been estimated with the Dicke formula in this work. Although this would, quite certainly, not alter our conclusions, this might be important when dealing with more favorable physical situations.
Obviously, a detailed time analysis of the signal would also be important and might slightly change the picture. This should be considered in the future. However, once again, this study was devoted to the investigation of the way axion haloscopes could be directly reused to search for gravitational waves.\\

In addition, it is worth pointing out that our analysis is {\it stricto sensu} valid only for masses quite below the upper bound, that is for systems seen long before the merging. Not only because post-Newtonian corrections are ignored but also because, even at the Newtonian level, it is not correct to assume, as we did, that the orbit remains circular -- albeit with a time-dependent radius -- close to the merging. The fact is that the trajectory is not a circle anymore just before the coalescence: there is {\it no} circle even approximately tangent to the trajectory. We do not expect any correction from the detailed shape of the orbit to drastically change our conclusions (the Fourier transform of the signal should even be wider, making the situation even worse) but, in principle, the present analysis is fully reliable only when $\mathcal{M}_C\ll \mathcal{M}_C^{max}$ (the latter corresponding to a gravitational wave frequency at merging equal to the cavity resonant one).\\

To summarize, the main new inputs of this work -- when compared with previous studies -- are the following. First, the frequency of the experiment does not fix the mass of the binary systems that could, in principle, be probed: a double degeneracy is at play. Second, and more importantly, the duration of the physical signal in the bandwidth of the instrument can be extremely small and taking this into account drastically reduces some of the usually assumed sensitivities, hence the number of expected events. This is especially relevant for systems generating the higher strain. Third, the combined effect of the noise averaging associated with the limited sampling rate with the partial cavity charging leads to a decrease of the effective quality factor reducing further -- by many orders of magnitude -- the sensitivity. Interestingly, when all the experimental effects are taken into account, the maximum probed distance does {\it not} depend anymore on the mass above $10^{-12}$ solar mass.

The numerical values we have obtained seem to close the window on compact binary systems as sources of GHz gravitational waves detectable by resonant cavities.\\

Beyond this somehow disapointing conclusion, we have tried to make this study as pedagogical as possible so that the methodology can be easily re-used for other setups.

\section{Acknowledgements}

JGB acknowledges support from the Research Project PID2021-123012NB-C43 [MICINN-FEDER], and the Centro de Excelencia Severo Ochoa Program CEX2020-001007-S at IFT.
TG acknowledges support from the French National Research Agency (ANR) in the framework of the GrAHal project (ANR-22-CE31-0025).

\section{APPENDIX}

In this short appendix, we further discuss the effective time entering the Dicke formula for the SNR (section III-A). More specifically, we focus on the case when $t_{\Delta\nu} < t_{min}$. Using heuristic arguments, we concluded  in subsection 3 that the mean power generated in the cavity includes a factor $\frac{t_{\Delta\nu}^2}{t_{min}^2}$, hence the effective time $\frac{t_{\Delta\nu}^4}{t_{min}^3}$. \\
This result can be also be foreseen in the following way. The cavity responds to an electromagnetic source like a lightly damped harmonic oscillator. The signal generated is thus the convolution of the source with the cavity impulse response, which is a free damped oscillation. At times shorter than the decay time $\frac{\omega}{2Q} \sim t_{min}$, the signal amplitude rises linearly with time: it grows like the number of source periods coherently superposed to the impulse response periods when performing the convolution. After the source has stopped, the signal decays. The overall response amplitude to a short source is thus proportional to its duration, and the energy and mean power are proportional to its square, that is to $t_{\Delta \nu}^2$  in our case.\\
It should however be also pointed out that considering the frequency modulated source to act only during $t_{\Delta \nu}$  is a conservative approximation. 
In practice, the frequency modulation only creates a small phase-shift, during the considered time interval, relative to the impulse response. Hence, the source is active over a longer time and we may anticipate that the transition between the intermediate and large mass regimes is actually less abrupt than described here.\\
A precise quantitative description would require a numerical study which is beyond the scope of the present paper and will be presented elsewhere. Although 
the resulting sensitivity should be slightly better, it will not alter our main conclusion about the unlikeliness of detecting black hole mergers using microwave cavities.

\bibliography{refs.bib}

 \end{document}